\RequirePackage{fix-cm}
\documentclass[twocolumn]{svjour3}          % twocolumn,referee
\smartqed  % flush right qed marks, e.g. at end of proof
\usepackage{graphicx}
\usepackage{newtxtext,newtxmath}

\usepackage[utf8x]{inputenc}
\usepackage[T1]{fontenc}
\usepackage{xcolor}

\usepackage{amsmath}

\usepackage{amsfonts}
\usepackage{lineno}
\usepackage[breaklinks,hidelinks]{hyperref}
\usepackage{ragged2e}
\usepackage{lineno}
\usepackage[UKenglish]{babel}

\newcommand{\edt}[1]{{\color{red}{#1}}}
\newcommand{\edtg}[1]{{\color{blue}{#1}}}
\DeclareMathOperator{\sign}{sign}

\usepackage[sort]{natbib}

\begin{document}

%\linenumbers

\title{The Middle Pleistocene Transition by frequency locking and slow ramping of internal period}



\author{Karl H.M. Nyman         \and
        Peter D. Ditlevsen %etc.
}


\institute{Karl H.M. Nyman \at
              Centre for Ice and Climate, Niels Bohr Institute, University of Copenhagen, Copenhagen, Denmark  \\
              \email{karl.nyman@nbi.ku.dk}
           \and
           Peter D. Ditlevsen \at
              Centre for Ice and Climate, Niels Bohr Institute, University of Copenhagen, Copenhagen, Denmark  \\
              \email{pditlev@nbi.ku.dk}
}

\date{Received: date / Accepted: date}

\maketitle

\begin{abstract}

The increase in glacial cycle length from approximately $41$ to on average $100$ thousand years around $1$ million years ago, called the Middle Pleistocene Transition (MPT), lacks a conclusive explanation. \edt{We describe a dynamical mechanism which we call Ramping with Frequency Locking (RFL), that explains the transition by an interaction between the internal period of a self-sustained oscillator and forcing that contains periodic components. This mechanism naturally explains the abrupt increase in cycle length from approximately $40$ to $80$ thousand years observed in proxy data, unlike some previously proposed mechanisms for the MPT. A rapid increase in durations can be produced by a rapid change in an external parameter, but this assumes rather than explains the abruptness. In contrast, models relying on frequency locking can produce a rapid change in durations assuming only a slow change in an external parameter. We propose a scheme for detecting RFL in complex, computationally expensive models, and motivate the search for climate variables that can gradually increase the internal period of the glacial cycles.}

\keywords{Glacial cycles \and Middle Pleistocene Transition \and Frequency locking \and Internal period \and Abrupt transition}
\end{abstract}

\section{Introduction}

\begin{figure*}
\begin{centering}
\includegraphics[width = 0.95\textwidth]{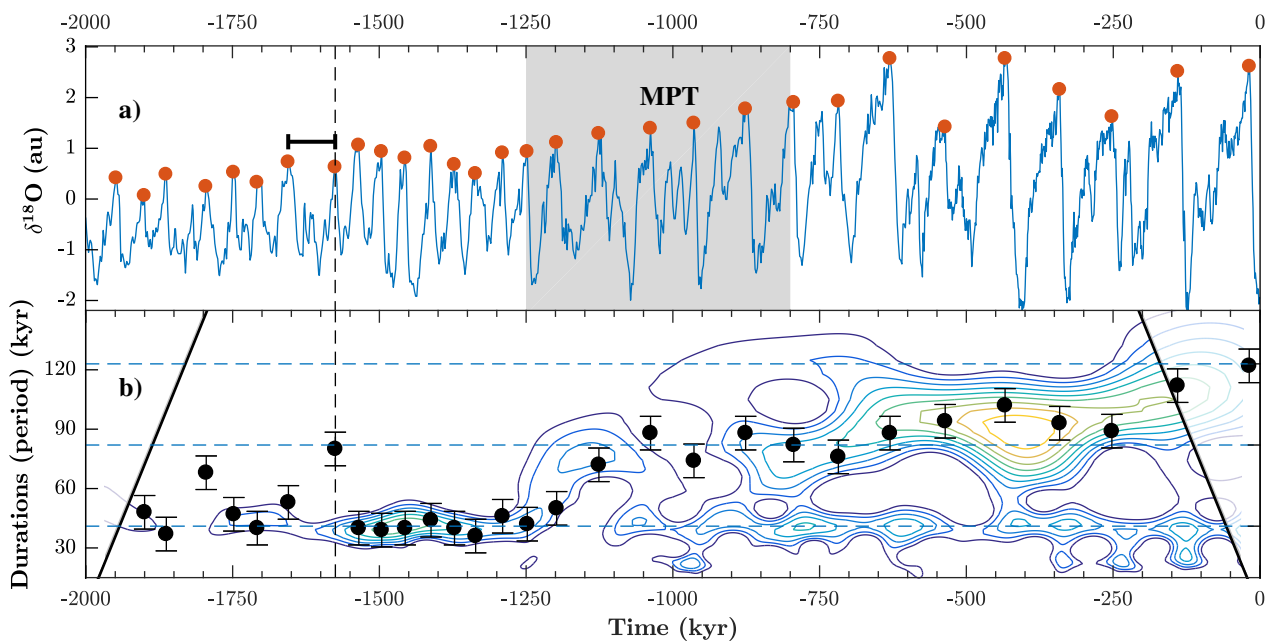}
\caption{a) The LR04 stack of normalised isotopic oxygen anomalies in deep ocean sediment cores --- a proxy for global ice volume and deep ocean temperatures \citep{lisiecki05}. Higher values means more ice. b) Durations between successive major terminations (black), showing an increase at the Middle Pleistocene Transition (MPT). Error bars indicate one standard deviation dating uncertainty. Contours (darker to lighter) show the amplitude of a wavelet spectral estimate. Each lighter contour corresponds to an increase of $7\%$ of the maximum amplitude, starting at $30\%$. Outside the cone of influence (thick black line), edge effects are important. See Appendix \ref{sec:wavelets} and \ref{sec:glacialterminations} for details}
\label{fig:benthicdata}
\end{centering}
\end{figure*}

Since the beginning of major Northern hemisphere glaciation $2.7$ million years (Myr) ago, Earth has undergone alternating epochs of icy and cold conditions on the one hand, and warm and ice-free conditions on the other (Fig. \ref{fig:benthicdata}). While these glacial cycles are attested from geological records \citep{lisiecki05, huybers07, epica04}, there is no single conclusive theory of their origins. Historically, the focus has been to explain the approximately $100$ thousand years (kyr) long glacial cycles that dominate the past $800$ kyr (see \citep{imbrie79} for a review). But as sediment records later revealed that these cycles were ca $40$ kyr long prior to $1.2$ Myr ago the question arose what caused the shift to approximately $100$ kyr long cycles, called the Middle Pleistocene Transition (MPT) \citep{clark06}.

The main strategy to address these questions has been to replicate the palaeoclimatic records using simple models of glacial cycles with few variables, referred to as conceptual models (see \citep{crucifix12} for a review). One reason for this is that the rather regular and cyclic variations in data suggest that the the main dynamics can be captured by a system of few degrees of freedom, even as the full climate system obviously has a large number of degrees of freedom. None of these models describe the climate system in detail, but they are useful for understanding underlying dynamics. Virtually all models involve insolation variations due to changes to Earth's orbital configuration relative to the sun, an idea heralded by Adhémar, Croll and Milankovitch \citep{imbrie79}. But the specific role played by insolation variations is still unknown and debated.

Several solutions to the cause of the MPT have been presented within the context of conceptual models. Some mechanisms rely on a bifurcation occurring in the unforced climate system which fundamentally changes how the system operates \citep{ashwin15,ditlevsen09,tziperman03,maasch90,huybers17}. Other mechanisms invoke a ``spontaneous'' change, such as a shift between attractors due to subtle changes in insolation \citep{quinn18,omta16} or random fluctuations \citep{salzman93, imbrie11}, or as a coincidence \citep{huybers09}. A third possible mechanism for the MPT assumes one essential mode of oscillation throughout the Pleistocene and relies crucially on the interaction between insolation variations and an increasing internal period. This mechanism, previously imprecisely referred to as phase/frequency locking and non-linear resonance --- but here Ramping with Frequency Locking (RFL) --- is the focus of this paper.

The main appeal of this mechanism is that nothing special had to occur in the climate system over the MPT \citep{huybers07}; it is only required that the internal period was ramped slowly --- interactions with forcing are enough to cause an abrupt increase in durations between glacial terminations (Fig. \ref{fig:benthicdata}, bottom panel).

The first publications where RFL was used \citep{paillard98,paillard04} did not explain why the durations between glacial transitions increased abruptly over the MPT. \citet{ashkenazy04,ashkenazy06} hinted how frequency locking (therein called phase locking) could produce an abrupt increase in duration, by showing diagrams of average duration as a function of a system parameter (Devil's staircases). \citet{huybers07} was first to both show a model trajectory of ice volume using the mechanism, and to attribute the effect to ``skipping of obliquity cycles'', a frequency locking effect. Recently, \citet{feng15,mitsui15,daruka15,tzedakis17} alluded to the mechanism, but neither emphasised that frequency locking can explain the MPT assuming only a slow linear change in a climate parameter. Instead, by ramping some parameter in a way that mimics the rapid change in durations over the MPT, they prescribe an abrupt increase in period over the MPT rather than explaining it. Here, we for the first time properly define RFL and emphasise its generality.

Following \citep{huybers07}, we question the common assumption that climate entered a stationary state in the late Pleistocene, and instead argue that the sequence of durations between glacial terminations is consistent with a slow increase of the internal period of the climate system until present (Fig. \ref{fig:benthicdata} b)). According to this view, the typical durations between transitions changed from ${\sim}40$ kyr to  ${\sim}80$ kyr around $1200$ kyr ago, after which they increased gradually in the mean to present time, with the last duration being ${\sim}120$ kyr long.

Here, we first aim to explain RFL in a clear way, using a harmonically forced simple model. We use harmonic (pure sine) forcing because it makes frequency locking concepts clearer, while still producing qualitatively similar behaviour to astronomical forcing curves. We should not expect model runs with such simplified forcing to agree well with data, however. We use forcing with period $41$ kyr, corresponding to the main period of obliquity variations \citep{berger78}, which determine the total insolation integrated over the summer at Northern latitudes \citep{huybers06}.

\edt{
We then define RFL, specify a class of models able to reproduce the MPT using the mechanism, and propose a decomposition of model components to understand the abruptness of the MPT. We consider evidence in data for a $40$ to $80$ kyr shift in durations between terminations and a subsequent gradual increase, and why this supports RFL in favour of some other mechanisms for the MPT. We then discuss how insights from harmonic forcing relate to non-harmonic forcing, how RFL can be detected in complex and computationally expensive models, and some climate variables that can cause an increase in the internal period of the glacial cycles.
}

\section{The idea behind Ramping with Frequency Locking}
\label{sec:idearfl}

\begin{figure}
\begin{centering}
\includegraphics[width = 0.48\textwidth]{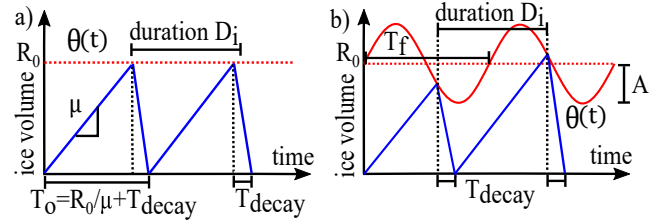}
\caption{The H07 model (Eq. \eqref{eq:H07}), a) unforced and b) periodically forced. Ice volume (blue) grows linearly at a rate $\mu$ until a threshold (red) is hit, after which ice volume is reset to $0$ over a time $T_{decay}$. In a) the threshold of glacial termination $\theta(t)$ is constant $\theta(t)=R_0$, whereas in b) it oscillates periodically as $\theta(t)=R_0 + A\sin{(2\pi t/T_f)}$. $T_o$ is the internal (unforced) period of the model}
\label{fig:modeldescription}
\end{centering}
\end{figure}

We illustrate Ramping with Frequency Locking (RFL) using a deterministic and continuous time version of the H07 model \citep{huybers07} (see Fig. \ref{fig:modeldescription}). \edt{The model is arguably the simplest to represent alternating stages of intrinsic growth and decay of ice sheets, with the growth state ending abruptly as a critical ice volume is reached. It is is an integrate-and-fire threshold model conceptually very similar to the models in \citep{vanderpol27,imbrie80,paillard98,ashkenazy04,huybers07,desaedeleer13,imbrie11,parrenin03,parrenin12,glass79}.

Physically, sudden and rapid deglaciation has been explained e.g. with isostatic rebound \cite{oerlemans80}, rapid $CO_2$ outgassing \citep{paillard04} and rapid loss of Northern hemisphere sea ice cover \citep{gildor00}.
}

We assume that ice volume $x(t)$ grows at a constant rate $\mu$ \edt{in a glacial state} until it reaches a threshold $\theta(t)$. \edt{Then deglaciation starts, whereby ice volume decays to $0$ over a fixed time $T_{decay}=10$ kyr:}
\begin{linenomath*}
\begin{equation}
\begin{aligned}
\dot{x} & = \mu \text{ until } x(t) = \theta(t) \text{, then} \\
& \text{ linearly decrease } x(t) \text{ to } 0 \text{ over \edt{time }} T_{decay}, \text{ repeat}.
\label{eq:H07}
\end{aligned}
\end{equation}
\end{linenomath*}
Small perturbations to the model, such as having a constant rate of decay instead of a fixed time, does not qualitatively affect its behaviour.

\edt{We split $\theta(t)$ into a \emph{forcing} term $A\cdot F(t)$ -- a zero-mean sum of periodic components -- and a \emph{ramping} term $R(t)$: $\theta(t)=R(t) + A \cdot F(t)$.}

\begin{figure*}
\begin{centering}
\includegraphics[width = 0.95\textwidth]{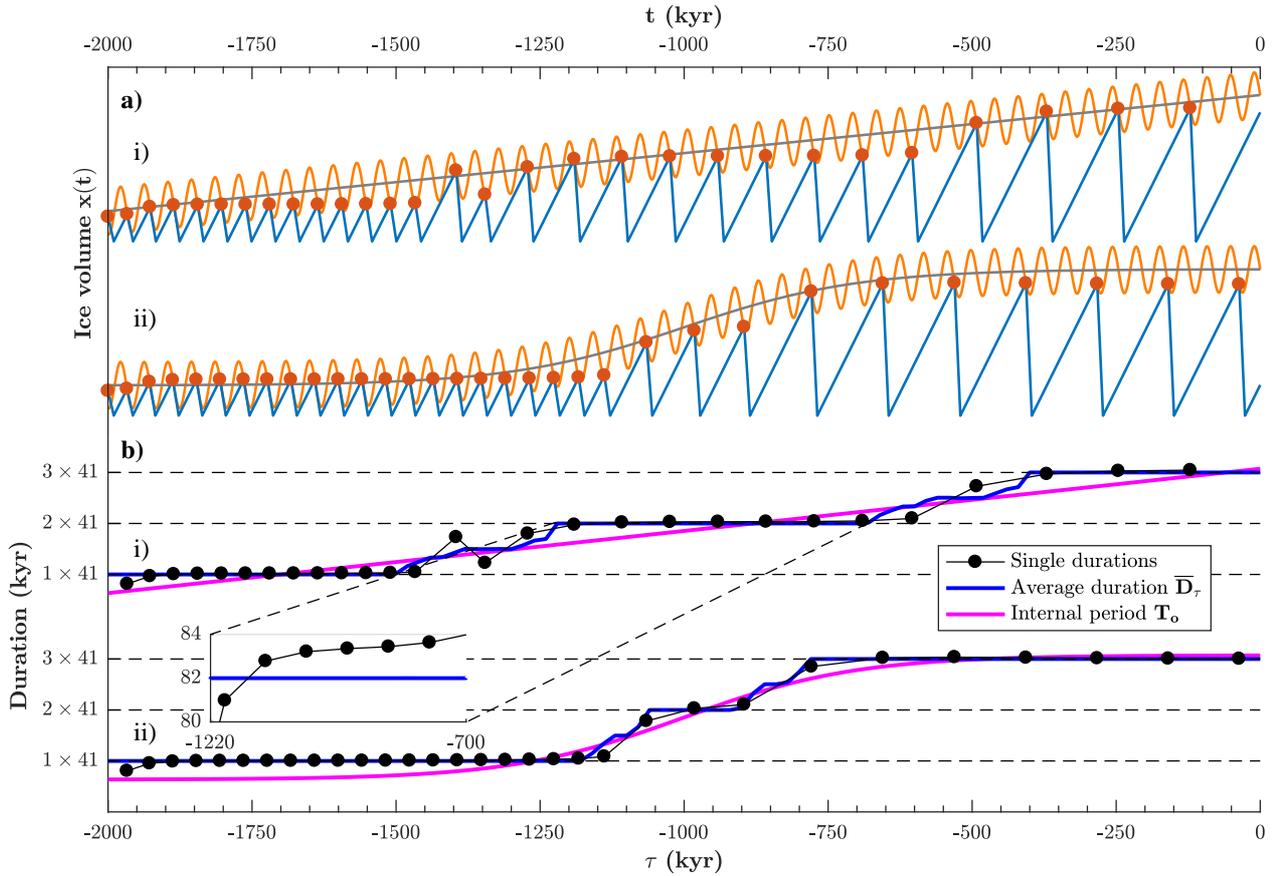}
\caption{The Ramping with Frequency Locking (RFL) mechanism for the periodically forced H07 model. a) Ice volume (blue sawtooth) over time for a i) linear and ii) sigmoidal ramp of the upper threshold $\theta(t)$. Periodic threshold in orange and unforced threshold in grey. b) Average duration between glacial terminations $\overline{D}_\tau$ over frozen time $\tau$ (black solid) (known as Devil's staircases, Sect. \ref{sec:arnoldtonguesdevilsstaircases}) for the i) linearly and ii) sigmoidally ramped thresholds, and sample durations (red dotted lines) for the forced solutions in a). Magenta lines show the internal period $T_o(\tau)$. The inset shows that durations with time-varying ramping $R(t)$ do not agree perfectly with the average duration $\overline{D}_\tau$, computed for $R(t)=\text{const}$ (see Sect. \ref{sec:quasistatic}). Parameters are $\mu = 1,T_{decay}=10, T_f = 41 ,A=20$ and ramp functions are i) $R(\tau) = 26 + 0.05\times (\tau + 2000)$ and ii) $R(t)=26+ 50(\tanh{(\frac{t + 1000}{300}) + 1)}$ respectively}
\label{fig:solutionsavgdurations}
\end{centering}
\end{figure*}

\edt{In the limit of constant ramping $R(t)=R_0$ and zero forcing $A=0$}, the system has a constant \emph{internal} period of oscillation $T_o = \frac{R_0}{\mu} + T_{decay}$ (subscript $o$ for \emph{oscillator}), \edt{see Fig. \ref{fig:modeldescription} a)}. \edt{But if the threshold increases slowly over time, for instance linearly $\theta(t)=R(t)=R_0 + R_1 t$ as in Fig. \ref{fig:solutionsavgdurations} a)i)}, then the internal period $T_o(t) = \frac{R_0}{\mu} + T_{decay} + \frac{R_1}{\mu}t$ also increases slowly (Fig. \ref{fig:solutionsavgdurations} b)i)). \edt{As there is no forcing, the durations between glacial terminations follow $T_o(t)$ closely.}

\edt{However, with periodic forcing \newline $\theta(t)=R_0 + R_1 t + A\sin{(2\pi/T_f)}$, durations $D_i$ are near multiples of the forcing period $D_i \approx N T_f$, $N \in \mathbb{N}$ (Fig. \ref{fig:solutionsavgdurations} b)i))}. Roughly speaking, the multiple that is realised is the one closest to the internal period $T_o$. \edt{This phenomenon, called frequency locking \citep{pikovsky01}, has been studied extensively over the past century (e.g. \cite{vanderpol27,cartwright45,glass79,letreut83,tziperman06})}.

In Fig \ref{fig:solutionsavgdurations} a)i) the durations $D_i$ change abruptly from $1\times T_f$ to $2\times T_f$ and finally $3\times T_f$. These abrupt changes in durations resulting from a gradual change in an underlying parameter is one possible dynamical mechanism behind the MPT.

We call the mechanism Ramping with Frequency Locking (RFL), rather than non-linear resonance, phase locking or frequency locking \edt{as it has previously been called. This we do to emphasise both that an internal period must increase gradually over time (ramping), and that the internal oscillations must be locked to external forcing}. This is opposed to e.g. the mechanism in \citep{omta16}, which \edt{realises the MPT through} jumps between coexisting frequency locked solutions.

\edt{We note that RFL is a special case of ``slow passage through bifurcation'' (e.g. \citep{do12,baer89}), for which the bifurcations typically are saddle-node bifurcations of limit cycles marking transitions in and out of frequency locking regions \citep{pikovsky01}. (However, see e.g. \cite{guckenheimer03,levi90} for other relevant bifurcations).

Finally, we note that H07 is an illustrative example of RFL, and not representative of all glacial cycle models.  However, the rapid jumps between frequency locking regions occur generically in a broad class of models, defined next.
}

\section{A formal description of RFL}
\label{eq:RFLdescription}

\edt{
H07 (Fig. \ref{fig:modeldescription}) is just one particular model capable of realising the MPT through RFL. \edt{We could simply call these self-sustained oscillators, but we aim to be more precise and to establish notation.}

First, we naturally require the model to be a dynamical system, such that there is an evolution rule $f(t,x)$ taking a state $x(t)$ forward in time $t$. We identify the model with the evolution rule and denote it $f$ (without arguments) for brevity. $f(t,x),x(t)$ and $t$ can be very general, for instance; $t$ can be continuous or discrete, $x$ can be of any dimension, and $f(t,x)$ can e.g. be a piecewise smooth ODE paired with a switching rule, as for H07.

We also require $f$ to be forced by a continuous zero-mean sum of periodic components $A(t) F(t)$ with an amplitude $A(t)$, called the \emph{forcing}. We further require that $f$ is parametrised by a set of parameters $p(t)$, whose time-varying subset $R(t)$ is called the \emph{ramping}. Thus we can write $f=f(t,x, R(t),A(t) F(t))$.

We define the \emph{frozen system} $f_\tau$$:= f(t,x,R(\tau),A(\tau)F(t))$ as $f$ with parameters frozen at time $t=\tau$. Importantly, we require that $f$ is a \emph{self-sustained} oscillator with internal period $T_o(R(\tau))$, meaning that every solution to $f_\tau$ with $A(t)\equiv0$ tends \emph{asymptotically} (as $t\to \infty$) to a periodic solution with period $T_o(R(\tau))$. For RFL to be relevant we require that $T_o(R(\tau))$ increases as a function of $\tau$. This is the \emph{ramping} part of RFL.

The \emph{frequency locking} part of RFL comes from the response of $f$ to non-zero but constant forcing $A(\tau)$. For small and medium size $A(\tau)$, asymptotic solutions to the \emph{frozen system} $f(t,x,R(\tau), A(\tau) F(t))$, are generically periodic with periods related rationally to the forcing periods \citep{pikovsky01}. The oscillator period can for instance be twice that of the forcing period. If so, the oscillator period (and therefore frequency) remains constant on open sets of parameters and we say that solutions are frequency locked to the forcing (we return to this in Section \ref{sec:breakdown}).

The essence of RFL is that the period of the frozen system can change rapidly as function of $T_o(R)$ when a ramped parameter causes the system to switch between frequency locking regions.

However, some remarks are in place. Firstly, the system with time varying parameters $f(t,x(t),R(t),A(t)F(t))$ is \emph{not} the same as the frozen system $f(t,x(t),R(t),A(\tau)F(\tau))$ since solutions to the former cannot equilibrate to solutions of the latter in finite time. We return to differences between the two systems in Section \ref{sec:quasistatic} but until then we focus on the frozen system.

Secondly, the period of an oscillator is not the same as the length of individual ``cycles''. For instance, around $-1350$ kyr in Fig. \ref{fig:solutionsavgdurations}, short and long ``cycles'' alternate. This makes the average time between terminations $61.5$ kyr, whereas the period (time until repetition, two large peaks) is $123$ kyr. Therefore, we instead characterise local behaviour with the \emph{average duration}
\begin{linenomath*}
\begin{equation}
\overline{D}_\tau = \lim_{n \to \infty} \frac{1}{n}\sum_{i = 1}^n D_{i,\tau},
\label{eq:avgperiodthreshold}
\end{equation}
\end{linenomath*}
where $D_{i,\tau}$ denotes the $i$:th duration between successive crossings of a fixed threshold for the frozen system $f_\tau$.

For some models like H07 glacial terminations are natural such thresholds. For other models Poincaré sections can be considered instead \citep{pikovsky01}.
}

\section{Breaking down the dependency of \texorpdfstring{$\overline{D}_\tau$}{Dtau} on \texorpdfstring{$\tau$}{tau}}
\label{sec:breakdown}

\edt{
Comparing Fig. \ref{fig:solutionsavgdurations} b)i) and b)ii) shows that $\overline{D}_\tau$ can rise steeply both from frequency locking effects under a gradual change of parameter (Fig. \ref{fig:solutionsavgdurations} b)i)) and from ramping of a climate parameter rapidly (Fig. \ref{fig:solutionsavgdurations} b)ii)).

We wish to break down the contribution to the local change in average duration from these effects and do so by considering the change $\Delta \overline{D}_\tau$ under a small perturbation $\Delta \tau$:
\begin{linenomath*}
\begin{equation}
\begin{aligned}
\frac{\Delta \overline{D}_\tau(\tau)}{\Delta \tau} \approx & \frac{\overline{D}_\tau(\tau + \Delta \tau) - \overline{D}_\tau(\tau)}{\Delta \tau} \approx \\
& \frac{\Delta \overline{D}_\tau(T_o,A)}{\Delta T_o}\frac{\Delta T_o(R)}{\Delta R} \frac{\Delta R(\tau)}{\Delta \tau} + \\
& \frac{\Delta \overline{D}_\tau(T_o,A)}{\Delta A} \frac{\Delta A(\tau)}{\Delta \tau},
\end{aligned}
\label{eq:abruptnessfunction}
\end{equation}
\end{linenomath*}
where e.g. $\frac{\Delta \overline{D}_\tau(T_o,A)}{\Delta T_o} := \frac{\overline{D}_\tau(T_o + \Delta T_o,A) - \overline{D}_\tau(T_o,A)}{\Delta T_o}$, and where we have neglected higher order terms. This approximation is generally better the smaller $\Delta \tau$ is. As $\Delta \tau \to 0$, \eqref{eq:abruptnessfunction} tends to the chain rule, but since $\frac{\Delta \overline{D}_\tau(T_o,A)}{\Delta T_o}=0$ wherever differentiable (see Section \ref{sec:arnoldtonguesdevilsstaircases}) it is more appropriate to consider $\Delta \overline{D}_\tau$ over short intervals of time $\Delta \tau$. In what follows we restrict ourselves to $\frac{\Delta A(\tau)}{\Delta \tau}=\nobreak0$.

Eq. \eqref{eq:abruptnessfunction} says that the rate of change (abruptness) in time of the average duration $\overline{D}_\tau$ is approximately the product of the rates at which $R(t)$ changes with time, $T_o$ changes with $R$, and $\overline{D}_\tau$ changes with $T_o$. Our point is that each of $\overline{D}_\tau$, $T_o$ and $R$ can contribute to an abrupt change of $\overline{D}_\tau$ at the MPT, but they have different interpretations from a modelling perspective. We discuss these factors next.
}

\subsection{\texorpdfstring{$\overline{D}_\tau(T_o,A)$}{Dtau(To,A)}: Arnold tongues and Devil's staircases}
\label{sec:arnoldtonguesdevilsstaircases}
The average duration $\overline{D}_\tau(T_o,A)$ as a function of internal period $T_o$ and forcing amplitude $A$ describes the frequency locking contribution to changes to $\overline{D}_\tau$ over time $\tau$.

\begin{figure}
\begin{centering}
\includegraphics[width = 0.48\textwidth]{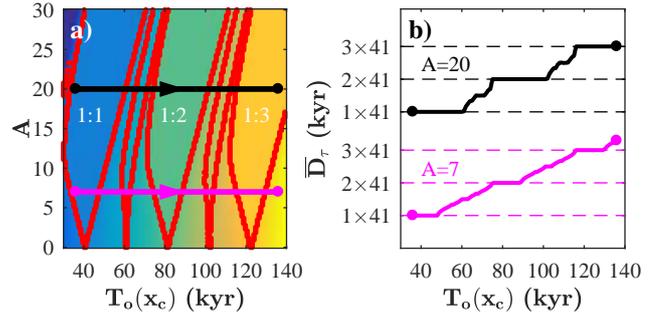}
\caption{a) Arnold tongue diagram for increasing upper threshold $R_0$ and periodic forcing ($T_f=41$ kyr) in the H07 model, showing regions in $(A,T_o)$ space of constant average duration $\overline{D}_\tau$ (enclosed by red dots). Major 1:N tongues, meaning that $\overline{D}_ \tau = N T_f$, are labelled. Colour scale from blue (short) to yellow (long) reflects average duration. b) Comparison between average duration $\overline{D}_\tau$ as a function of internal period $T_o$ for strong ($A=20$) and weaker ($A=7$) forcing}
\label{fig:arnoldtongues}
\end{centering}
\end{figure}

Frequency locking can be visualised in Arnold tongue diagrams; Fig. \ref{fig:arnoldtongues} a) reveals regions of constant average duration $\overline{D}_\tau$ in $(T_o, A)$ space called Arnold tongues \citep{pikovsky01,crucifix13,desaedeleer13}. Inside major 1:N tongues, solutions are periodic with period $N$ times the forcing period $T_f=\nobreak41$ kyr, as evidenced in (Fig. \ref{fig:arnoldtongues} a)). Minor tongues emanate at $A=0$ from other rationals of $T_f$, and in between them are quasiperiodic solutions. (We show only $M:N, ~M=\nobreak\{1,2\}$ Arnold tongues, defined numerically as sets for which $|\overline{D}_\tau - T_f\frac{N}{M}|<\nobreak0.5$. $\overline{D}_\tau$ is estimated over $6$ million years.)

A change in $\tau$ that in turn leads to a change in $T_o(R(\tau))$ traces out a path in $(A,T_o)$ space (black and magenta lines in Fig. \ref{fig:arnoldtongues} a). Such a path represents the change in system state as one or more parameters change in time over the MPT. The paths in Fig. \ref{fig:arnoldtongues} a) pass through the major 1:1, 1:2 and 1:3 locking tongues, in which there are respectively $1$, $2$ and $3$ forcing periods per oscillator period. We learn that for larger $A$, a larger portion of the path stays inside the major 1:N tongues, an observation also made in \citep{ashkenazy06}.

Another way of visualising the change in average duration $\overline{D}_\tau$ as a function of $T_o$ are Devil's staircases \citep{pikovsky01} (Fig. \ref{fig:arnoldtongues} b)), in which the forcing amplitude $A$ is fixed. We see that the average duration $\overline{D}_\tau$ is constant within Arnold tongues and that the staircase for larger $A$ contains longer steps of constant duration, as predicted from Fig. \ref{fig:arnoldtongues} a). Hence, stronger forcing influence tends to cause more abrupt changes to the average period.

\subsection{\texorpdfstring{$T_o(R)$}{To(R)} and \texorpdfstring{$R(\tau)$}{R(tau)}}

The function $T_o(R)$, if continuous and monotonic, stretches and squeezes Arnold tongues by scaling the independent variable $T_o$ of $\overline{D}_\tau(T_o)$. In the Ashke\-nazy model \citep{ashkenazy06}, for instance, a faster-than-linearly increasing $T_o(R)$ makes the 1:2 and 1:3 Arnold tongues, as a function of ice volume threshold, narrower and more closely spaced than the 1:1 tongue.

\begin{figure}
\begin{centering}
\includegraphics[width = 0.48\textwidth]{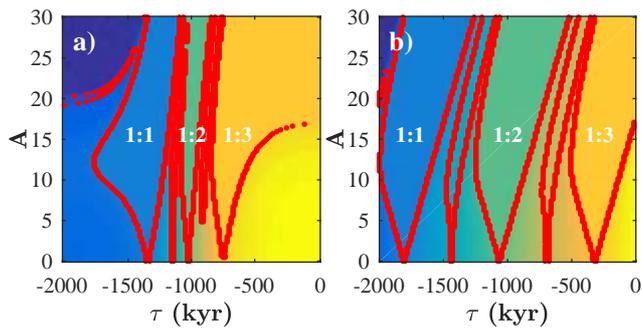}
\caption{Stretching of Arnold tongues by ramping the threshold parameter $R(\tau)$ in H07 at different rates. In a) $R(\tau)$ is ramped sigmoidally $R(\tau) = 20 + 55\times(\tanh( \frac{\tau + 1000}{300}) + 1)$, while in b) $R(\tau) = 20 + 0.055\times (\tau + 2000)$. For further details, see Fig. \ref{fig:arnoldtongues} and the text}
\label{fig:arnoldtonguesqueezecomparison}
\end{centering}
\end{figure}

Ramping $R(\tau)$ continuously and monotonically, similarly stretches and squeezes Arnold tongues. For instance, in Fig. \ref{fig:arnoldtonguesqueezecomparison} a sigmoidal ramping $R(t)$ makes the 1:2 Arnold tongue narrower compared to a linear change of $R(t)$. Fig. \ref{fig:solutionsavgdurations} further illustrates this, showing model runs for either a sigmoidally ($R(t) =\nobreak26 + 50\times \nobreak(\tanh( \frac{t + 1000}{300}) + 1)$) or a linearly $R(t) =\nobreak26 +\nobreak 0.05\times (t + 2000)$ ramped threshold. The sigmoidal ramping accelerates the increase in average duration around $-1000$ kyr, making the transition more abrupt. Note that the parameters in the functions $R(\tau)$ in Fig. \ref{fig:arnoldtonguesqueezecomparison} and Fig. \ref{fig:solutionsavgdurations} are different.

\subsection{The roles of \texorpdfstring{$\overline{D}_\tau(T_o,A)$}{Dtau(To,A)}, \texorpdfstring{$T_o(R)$}{To(R)} and \texorpdfstring{$R(\tau)$}{p(tau)} in reproducing the MPT}

All of $\overline{D}_\tau(T_o,A)$, $T_o(R)$ and $R(\tau)$ govern the average duration $\overline{D}_\tau$ and are able to cause an abrupt change of it, like the one observed at the MPT. From a modelling point of view, however, the functions carry different assumptions and are compatible with different hypotheses.

A model having an abrupt change due to $\overline{D}_\tau(T_o,A)$ relies on frequency locking properties, and assumes only slowly varying functions $T_o(R)$ and $R(\tau)$. Hence, the internal period is assumed to change slowly with model parameters, and parameters are assumed to change slowly in time. Such a model, relying on few assumptions about the climate system, makes full use of the RFL mechanism. The models in \citep{paillard98,paillard04,huybers07} and H07 are of this kind.

A model which relies predominantly on $T_o(R)$ for an abrupt change in average duration $\overline{D}_\tau$ is also consistent with a slowly changing external parameter $R(\tau)$, but a particular function $T_o(R)$ requires a physical explanation.

A model relying on a rapidly changed external parameter $R(\tau)$ does not need frequency locking properties of $\overline{D}_\tau(T_o,A)$ or a non-linear response of internal dynamics to the parameter $T_o(R)$. However, such a model prescribes the abrupt change in average duration at the MPT rather than explaining the dynamics behind it. Therefore, such an explanation requires justification for the rapidly changed external parameter. The models in  \citep{tzedakis17,mitsui15,ashkenazy04,daruka15} can be said to fall under this category, although they also achieve some abruptness through $\overline{D}_\tau(T_o,A)$.

\section{Validity of the quasi-static approximation \texorpdfstring{$f \sim f_\tau$}{f approx ftau}}
\label{sec:quasistatic}

\edt{
The quasistatic approximation is the approximation that parameters $R(\tau)$ change so slowly that the local average duration of $f$ at time $t=\nobreak\tau$
\begin{linenomath*}
\begin{equation}
\overline{D}_{\tau,loc} = \frac{1}{|I(\tau)|}\sum_{i \in I(\tau)} D_{i,\tau},
\label{eq:locavgperiod}
\end{equation}
\end{linenomath*}
is equal to $\overline{D}_\tau$. $I(\tau)$ is the set of indices of durations $D_{i,\tau}$ within a time interval $[\tau -\tau_0,\tau + \tau_1]$ around $\tau$, with $\tau_0,\tau_1>0$. If $I(\tau)=\emptyset$, then we define $\overline{D}_{\tau,loc}=0$.}

If the quasistatic approximation holds, then the average duration, Arnold tongue diagrams and Devil's staircases calculated for the frozen system $f_\tau$ provide accurate information about local dynamics of $f$.

However, if $R(t)$ and/or $T_o(R(t))$ change rapidly around $t = \tau$, then there are two sources of discrepancy between $\overline{D}_{\tau,loc}$ and $\overline{D}_\tau$.

\begin{figure*}
\begin{centering}
\includegraphics[width = 0.95\textwidth]{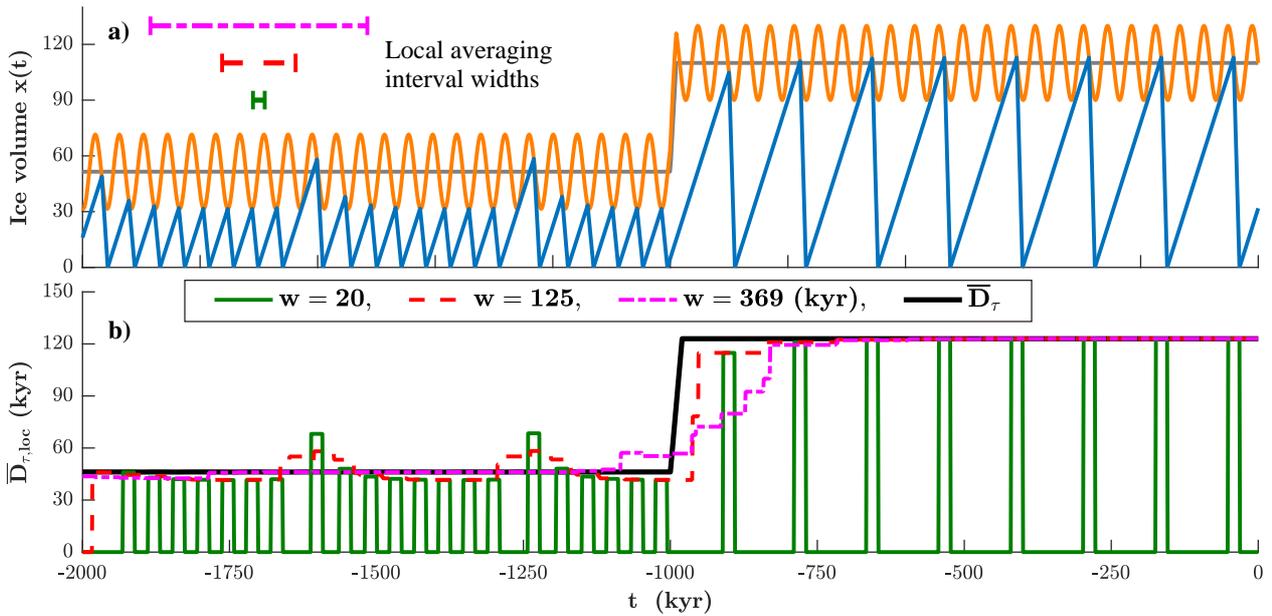}
\caption{Illustration of the difficulty estimating a local average frequency. a) Ice volume from H07 (blue sawtooth), threshold of glacial termination (orange) and mean threshold of glacial termination $R(t)$ (grey). b) shows estimates of the local average duration $\overline{D}_{\tau,loc}$ between glacial terminations in the H07 model with periodic forcing ($T_f=41$ kyr), for different window widths. b) $\overline{D}_{\tau,loc}$ (green solid, red and magenta dashed) are running averages of durations in sliding windows of width $w=[20,125,369]$ kyr, or $0$ kyr if there are no durations in a window. The frozen time average duration $\overline{D}_\tau$ (black solid) is shown for reference. Prior to $-990$ kyr the steady state solution has a period of $9\times 41=369$ kyr, but an average duration $\overline{D}_\tau = 46.125$ kyr. The mean threshold of glaciation $R(t)$ is ramped from $51.5$ to $110$ over $10$ kyr ($-990$ to $-1000$ kyr). Other parameters are $\mu=1,T_{decay}$ and $A=20$}
\label{fig:illustratedlocwidth}
\end{centering}
\end{figure*}

The first comes from that the length of the interval of time needed for a good average may be long relative to the local change of $\overline{D}_\tau$ for the system $f_\tau$. Fig. \ref{fig:illustratedlocwidth} illustrates that for a $9\times 41=\nobreak369$ kyr-periodic solution, a long interval is needed to get a local average duration $\overline{D}_{\tau,loc}$ in agreement with $\overline{D}_{\tau}$. At the same time, a long averaging interval fails to capture abrupt changes to $\overline{D}_\tau$.

The second is that solutions to $f$ may fail to track solutions to $f_\tau$. This occurs if the ``frozen'' attractor of $f_\tau$ changes (in some sense) at a fast rate, and if solutions attract to the frozen attractor at a slow rate. Quantifying these rates in a coordinate- and model-independent way seems difficult, however.

A candidate measure of rate of attraction is the maximal Lyapunov exponent of the return map mapping one transition time to another \citep{pikovsky01}. This can be normalised to a common time scale between models and is coordinate independent. However, since it is only a local measure it neglects the time it takes to enter a small neighbourhood of the attractor. This time can in practice dominate, as is the case in the standard circle model (not shown, model described in \citep{pikovsky01}).

The local change in average duration $|\frac{\Delta \overline{D}_\tau}{\Delta \tau}|$ is a candidate measure of rate of change of an attractor of $f_\tau$, since it exists in all models $f$ and is coordinate-independent. It is ambiguous how large $\Delta \tau$ should be, however. Furthermore, the average duration $\overline{D}_\tau$ is only a proxy for the position of an attractor in phase space; an attractor can move even if $\frac{\Delta \overline{D}_\tau}{\Delta \tau}=\nobreak0$. This explains the consistent deviation of single durations from the predicted and locally constant $\overline{D}_\tau=\nobreak82$kyr in the inset of Fig. \ref{fig:solutionsavgdurations} b) i).

\section{Is there a 100 kyr world?}

The late Pleistocene (${\sim}800$ -- $0$ kyr) is sometimes referred to as the ``$100$ kyr world'', carrying the implicit notion that the Earth system has settled in a stationary mode with a dominant time scale of $100$ kyr (Fig. \ref{fig:benthicdata} a)). This view, originating from the closeness to the $100$ kyr component of eccentricity (an astronomical parameter), is supported by the rate of increase of mean ice volume seemingly levelling off \citep{clark06,mudelsee97}, and that the Fourier spectrum over the last ${\sim}800$ kyr is centred around $100$ kyr.

We propose on the contrary, following \citep{huybers07}, that the glacial period increased gradually from ${\sim}80$ kyr around $-1200$ kyr to ${\sim}120$ kyr at present day. The change from ${\sim}40$ to ${\sim}80$ kyr long cycles at $-1200$ kyr can be a shift from $1\times41$ to $2\times41$ kyr obliquity frequency locking, and/or $2\times21$ to $4\times21$ kyr precession locking. We base this claim on durations between major glacial terminations and a wavelet spectrum of the LR04 stack (see Fig. \ref{fig:benthicdata} b)); both quantities increase rather rapidly around $-1200$ kyr and show a steady but irregular increase towards present time.

\edtg{
\subsection{Identifying the shift to longer periods}

While \citet{huybers07} observed that the mean period of global ice volume variations increases over time, we \edtg{make the stronger} claim that an abrupt shift from $40$ to $80$ kyr long durations occurred around $-1200$ kyr. We base this claim on our identification of major glacial terminations, which unlike spectral decomposition ignores glacial cycle shape and is unaffected by time-frequency resolution.

A disadvantage of using glacial termination events is that it is unclear what constitutes a major termination, and whether it is meaningful to characterise glacial cycles by termination events. Nevertheless, we believe that our identification of major terminations is sufficiently robust to support the claim that the duration shifted abruptly from ${\sim}40$ to ${\sim}80$ kyr around $-1200$ kyr.
}

\edtg{
\subsection{Testing for trend after the MPT}
\label{sec:testfortrend}

It appears that the durations between successive glacial terminations are increasing over time starting at the onset of the MPT around $-1200$ kyr.

We evaluate whether this trend is statistically significant, using a variation on the Mann-Kendall test \citep{mann45,kendall55}. Our null hypothesis H0 is that the sequence of thirteen durations from $-1126$ kyr until present is generated by a process with stationary mean, and that any observed monotonicity is by chance. Since $\tilde{D}_i = D_i - D_{mean}$, successive deviations from the mean duration $D_{mean}=91$ kyr are correlated, we immediately reject a white noise process as assumed in the standard Mann-Kendall test. Instead, we model them as an AR(1) process, such that $\tilde{D}_{i+1} = \alpha \tilde{D}_{i} + \sigma_d \xi_i$, where $\xi_i$ are independent Gaussian zero mean and unit variance elements. The parameters $\alpha=0.6$ and $\sigma_d=14.5$kyr are the standard estimates of lag 1 and 0 autocorrelation coefficients respectively.

We test the hypothesis using the Kendall $\tau$ test statistic for monotonicity $\tau_K$, based on the number of ordered and disordered pairs in a sequence. $\tau_K=1$ for a perfectly ordered sequence and $\tau_K=0$ for sequence with equally many ordered and disordered pairs (see Appendix \ref{sec:kendalltau} for a definition of $\tau_K$). We evaluate $\tau_K$ for $2\cdot10^4$ samples of the AR(1) process. As indicated in Fig. \ref{fig:kendalltau}, it is unlikely $(p<0.05)$ to observe the test statistic in durations from data, assuming that the durations follow an AR(1) process. Therefore, we reject the null hypothesis of no trend.

Adding age model uncertainty to the Monte Carlo sequences of durations only makes it more difficult to reject H0. Furthermore, slightly different choices of major glacial terminations, or the use of an untuned record, does not influence the conclusion of the test.

\begin{figure}
\begin{centering}
\includegraphics[width = 0.48\textwidth]{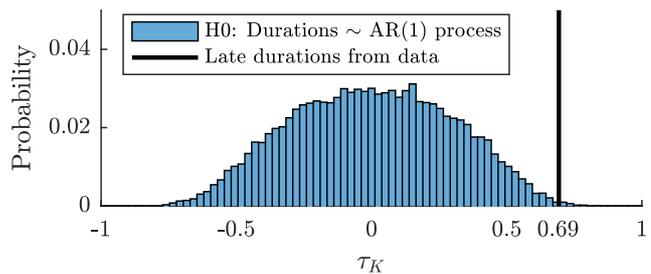}
\caption{Histogram shows a Monte Carlo distribution of the Kendall tau ($\tau_K$) test statistic, under the null hypothesis H0 that the sequence of durations between glacial terminations from $-1126$ kyr follow an AR(1) process. Larger $\tau_K$ indicates a more monotonic sequence. Black line shows the test statistic $\tau_K$ for durations in an ice volume proxy (Fig. \ref{fig:benthicdata}). $\tau_K$ under H0 exceeds the observed $\tau_K$ only in $5\%$ of the cases. For details, see Section \ref{sec:testfortrend}}
\label{fig:kendalltau}
\end{centering}
\end{figure}

\subsection{Consequences for modelling the MPT}

Some explanations for the MPT do not reproduce the sequence of successively longer durations between glacial terminations in data as naturally as RFL. Instead, they produce long period cycles at the onset of the MPT which shorten towards the present as a parameter is ramped.

The Maasch and Salzman 1990 model in Fig. \ref{fig:sm90} is one such model \citep{maasch90}. The inconsistency with data is evident when comparing the model durations with those in the LR04 stack (Fig. \ref{fig:benthicdata}). Another such model is the Tziperman and Gildor 2003 model \citep{tziperman03}.

Although different dynamical mechanisms are at play in these models, they have in common that a long period limit unforced cycle emerges near a region of slow motion in phase space. As a parameter is varied, the limit cycle moves farther from this region, shortening the internal period.

RFL on the other hand naturally explains both a sudden shift from $40$ kyr to $80$ kyr cycles and a gradual increase towards longer cycles, since the system can respond both smoothly and abruptly to an increasing internal period, due to the Devil's staircase structure (e.g. Fig. \ref{fig:arnoldtongues}). We interpret the progression of durations as evidence against models like Maasch and Salzman 1990 and Tziperman and Gildor 2003, and for mechanisms that naturally produce increasing glacial cycle length, such as RFL.

\begin{figure}
\begin{centering}
\includegraphics[width = 0.48\textwidth]{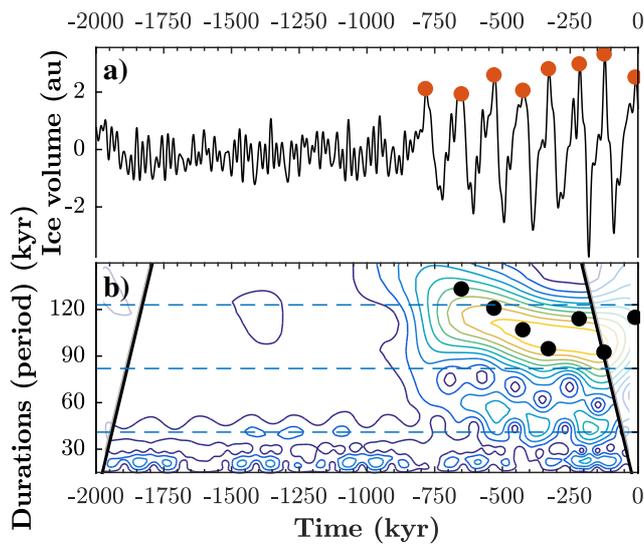}
\caption{Simulation of the Maasch and Salzman model in \citep{maasch90} forced by Summer solstice insolation at $65$ degrees North. a) Global ice volume over time (black), with glacial terminations (red dots) at peaks chosen for simplicity to be above $1.2$ normalised ice volume units and spaced at least $60$ kyr apart. Self-sustained cycles emerge around -$800$ kyr and shorten towards the present. b) Durations and wavelets as in Fig. \ref{fig:benthicdata}, except that contours start at $10\%$ of the maximum wavelet amplitude}
\label{fig:sm90}
\end{centering}
\end{figure}

}

\section{Non-harmonic forcing}

\edtg{
RFL is not restricted to harmonic forcing, but occurs also for astronomical, non-harmonic forcing. This is for instance the case for the Paillard and Parennin 2004 model (Fig. \ref{fig:pp04}), forced by summer solstice insolation at $65$ degrees North (65Nss, Fig. \ref{fig:65Nss} a)). As a parameter is increased linearly, durations first cluster around $41$ kyr, then shift abruptly to cluster around $82$ kyr at $-1000$ kyr, after which they increase gradually until present. The shift to $80$ kyr durations is later than in proxy data (Fig. \ref{fig:benthicdata}) and there are some short and long durations not clear in the proxy record, but overall the glacial terminations coincide well.

\begin{figure}
\begin{centering}
\includegraphics[width = 0.48\textwidth]{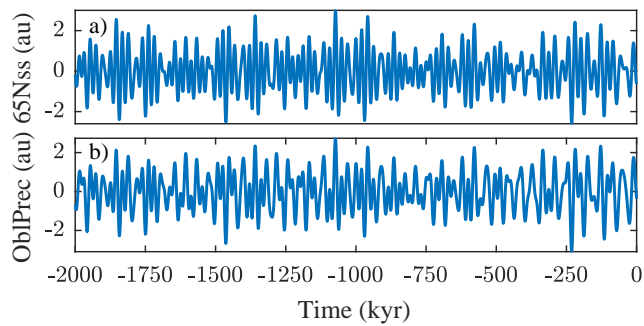}
\caption{Astronomical insolation curves. a) Summer solstice insolation at 65 degrees North (65Nss), normalised to zero mean and unit variance \citep{laskar04}. The signal is approximately a linear combination of $33\%$ normalised obliquity and $77\%$ normalised precession, two modulated sinusoidal signals with central frequencies $41$ and $22$ kyr \citep{crucifix13}. b) A normalised insolation curve consisting of $50\%$ obliquity and $50\%$ precession}
\label{fig:65Nss}
\end{centering}
\end{figure}

\begin{figure*}
\begin{centering}
\includegraphics[width = 0.95\textwidth]{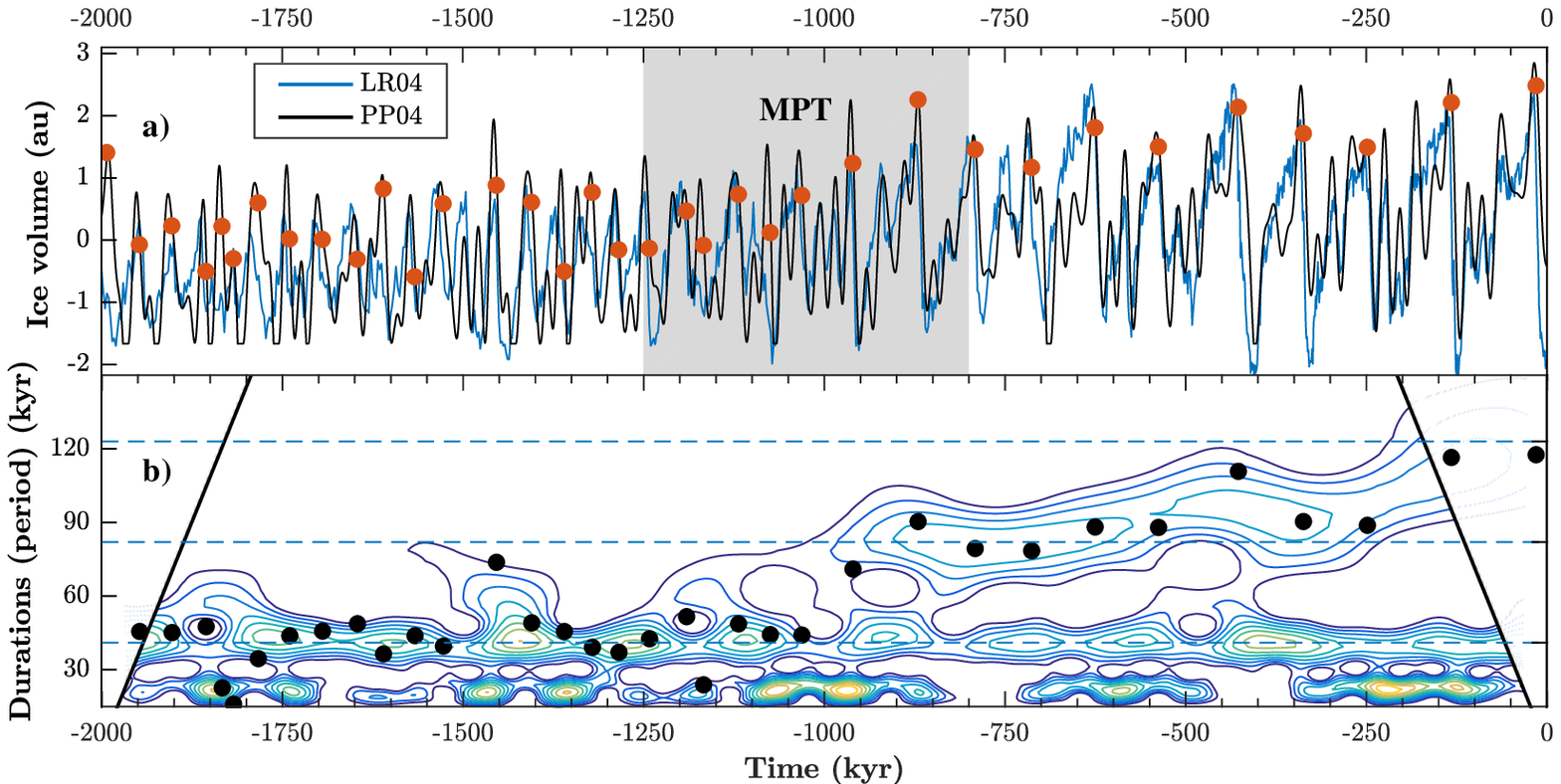}
\caption{Simulation of the Paillard and Parennin 2004 model in \citep{paillard04} forced by Summer solstice insolation at $65$ degrees North (Fig. \ref{fig:65Nss} a)). a) Model ice volume over time (black) contrasted with the LR04 stack (blue) (Fig. \ref{fig:benthicdata}), with glacial terminations (red dots) at times when a switch in Southern ocean circulation occurs. b) Durations and wavelets as in Fig. \ref{fig:benthicdata}}
\label{fig:pp04}
\end{centering}
\end{figure*}

Multi-frequency forcing generally produces Devil's staircases with shorter steps of constant duration, making them look ``smooth'' (e.g. Fig. \ref{fig:staircasecomparison}). This is apparently a problem for RFL since it relies on rapid jumps in durations. However, RFL can still be relevant as demonstrated by H07 forced by an equal amount of obliquity and precession (Fig. \ref{fig:65Nss} b), Fig. \ref{fig:H07equaloblprec} and Fig. \ref{fig:staircasecomparison}). Such forcing corresponds well to e.g. caloric summer insolation or integrated insolation above a threshold \citep{huybers11, tzedakis17}. The median and mode of the distribution of durations change more abruptly than the mean, which reflects that the gradual increase in average duration is caused by a gradual redistribution of durations between clusters, rather than a gradual increase of the most typical durations. In a simulation with time-dependent ramping parameter, the local-in-time distribution of durations cannot be sampled well. Therefore the majority of the realised durations come from the dominant clusters of durations, which can give the impression that durations shift rapidly, in spite of the average duration changing gradually (Fig. \ref{fig:H07equaloblprec} and Fig. \ref{fig:staircasecomparison})).

\begin{figure}
\begin{centering}
\includegraphics[width = 0.48\textwidth]{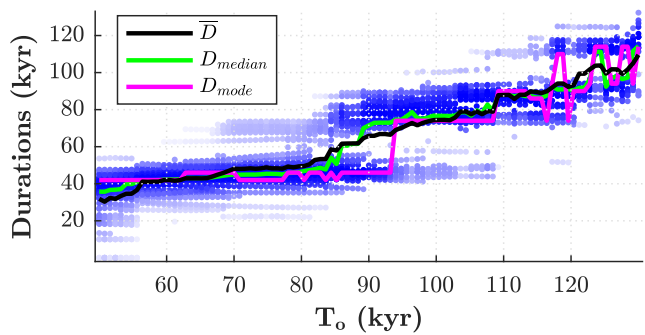}
\caption{Devil's staircases of H07 forced by equal amounts of obliquity and precession (Fig. \ref{fig:65Nss} b)). The average duration $\overline{D}$ (black line) as function of internal period $T_o$ is gradually increasing, whereas the the median $D_{median}$ and the mode $D_{mode}$ are more step-like. Blue dots are the population of durations for fixed internal period; darker colours indicate higher density of durations. $D_{mode}$ is defined from binning the durations; $D_{mode}$ is the mean of the edges of the $4$-kyr bin with the highest frequency. All quantities are evaluated from $-2000$ kyr to the present. Model parameters as in Fig. \ref{fig:H07equaloblprec}}
\label{fig:staircasecomparison}
\end{centering}
\end{figure}

\begin{figure*}
\begin{centering}
\includegraphics[width = 0.95\textwidth]{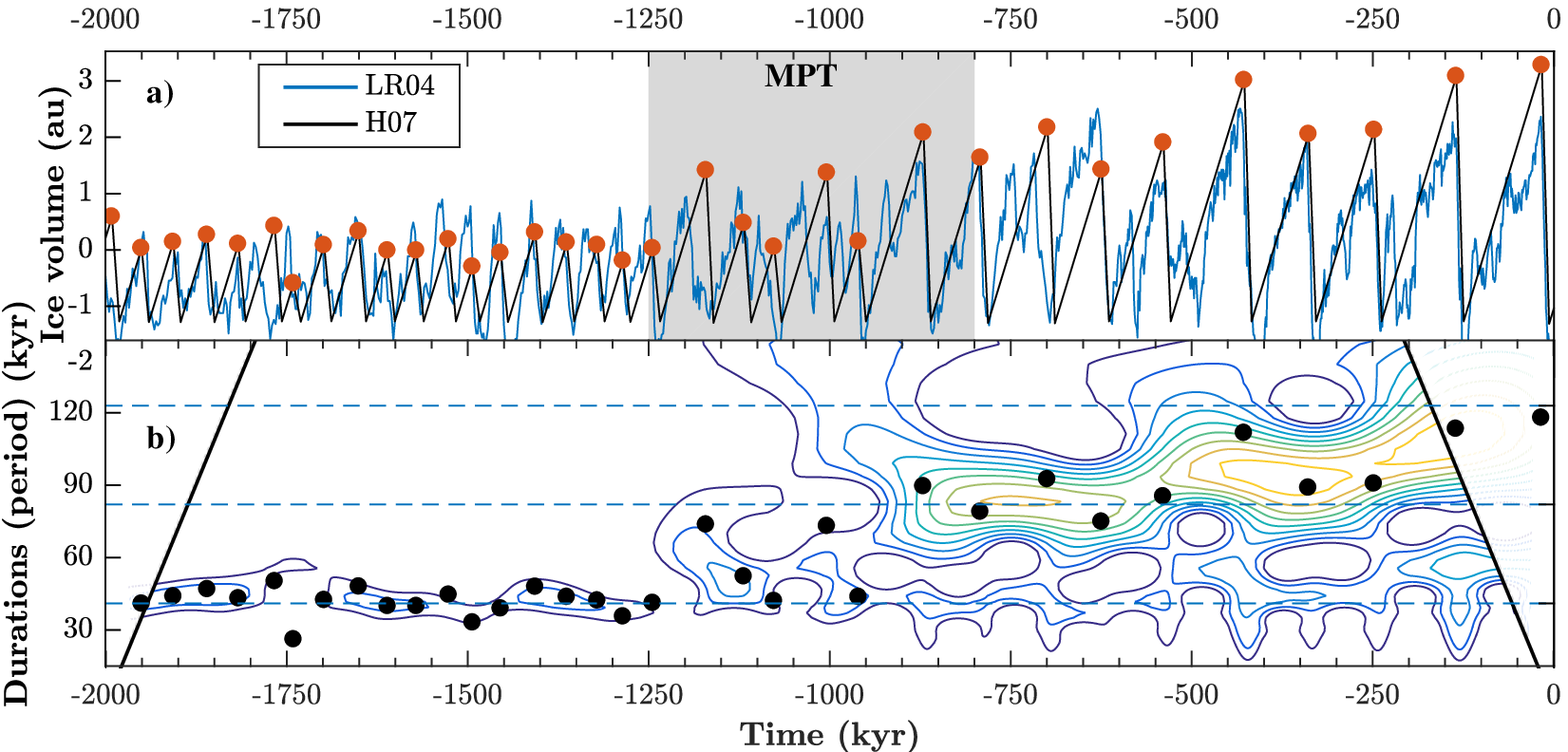}
\caption{Simulation of the H07 model forced by a sum of equal amounts obliquity and precession (Fig. \ref{fig:65Nss} b)). a) Model ice volume (black) shown with the LR04 stack (blue) (Fig. \ref{fig:benthicdata}), with glacial terminations (red dots) at times when a threshold of deglaciation is reached. b) Durations and wavelets as in Fig. \ref{fig:benthicdata}. The threshold of deglaciation increases linearly as $R(t)=40 + 0.04(t + 2000)$, and forcing amplitude is $A=26$. All other parameters are as in Section \ref{sec:idearfl}}
\label{fig:H07equaloblprec}
\end{centering}
\end{figure*}

Multi-frequency forcing gives rise to many interesting phenomena regarding predictability of solutions, see for instance \citep{tziperman06,crucifix13,grebogi84,mitsui15,desaedeleer13,letreut83,ashwin18,imbrie80}. Importantly, however, these phenomena are not essential to RFL. Whether solutions are truly frequency locked or depend on initial conditions is irrelevant, as long as durations undergo abrupt change and tend to cluster.

We conclude that RFL, clearly understood under periodic forcing, also is relevant for astronomical forcing. Indeed, recent studies provide evidence for the long-standing hypothesis that a combination of precession and obliquity paces the glacial cycles \citep{feng15,huybers11,tzedakis17}. Differences between periodic and multi-frequency forcing exist, but are not crucial for modelling the MPT with RFL.
}

\edtg{

\section{Relevance for complex models and physical mechanisms}

We see two practical uses of our description of RFL: To guide modelling of the MPT in complex models, and to drive the search for slowly changing climate variables.

\subsection{Relevance for complex models}

While climate physics are highly simplified in conceptual models like H07, their dynamics are well understood. The opposite holds true for Earth System Models (ESMs), which resolve multiple processes of climate in detail. To learn if the dynamical mechanism of RFL applies to such a model, we could in theory produce an Arnold tongue diagram as for H07 (Fig. \ref{fig:arnoldtongues}). However, since running ESMs is computationally expensive this is presently not possible. Nevertheless, it might be possible to detect signatures of RFL from only few model runs.

First, one should investigate whether the glacial cycles are self-sustained by fixing model parameters at plausible values and fixing the insolation field at its mean value. This is the case if, after a transient time, variations in ice volume on the order of $10$-$100$ kyr persist.

The next step is to sparsely sample an Arnold tongue diagram. First, a ramping parameter must be chosen. This does not have to be a scalar, but can be a function like a parametrisation, as long as its change over time is well defined. The parameter should be one that feasibly could influence the internal period of glacial cycles.

If changing the parameter changes the internal period, then one can compare the average duration of (insolation variation) forced and unforced solutions. If the average duration of the forced solutions is close to either 40 or 80 kyr and remains close even under parameter perturbations that change the internal period, then this is an indication that the system is frequency locked to insolation in a way relevant for the glacial cycles. In that case, there is good reason to research RFL more closely in the model.

\citet{ashkenazy06} suggested that synchronisation can be detected by running the system from multiple initial conditions and see if solutions converge. This procedure is not enough for us; we need to know if the internal period can be shifted appropriately with a change in parameter, and we need to know if the durations can robustly cluster on $40$ and $80$ kyr.

\subsection{Ramped climate variables}

To evaluate whether RFL caused the MPT one must identify slowly changing climate variables. Two such candidate variables are atmospheric $CO_2$ and atmospheric or oceanic temperatures. Since less $CO_2$ leads to a generally cooler atmosphere, it can be viewed as a proxy for global average atmospheric temperature. Local cooling can occur for other reasons, however.

There are curr\-ently no direct measure\-ments of atmo\-spheric $CO_2$ across the MPT, but a recent re\-con\-struc\-tion back to $-2000$ kyr suggests that the mean $CO_2$ did not change in the mean until at least $-1300$ kyr \citep{honisch09}. Since the reconstruction implies that $CO_2$ fell $31$ ppm by $-700$, the decrease in $CO_2$ must either have been rapid and driving the MPT, or a consequence of it. A rapid change in $CO_2$ is still consistent with RFL, but in that case RFL does not explain the abrupt increase in cycle length at the MPT, and instead one must find an explanation for the rapid increase in $CO_2$. However, the planned European BEOIC deep ice core drilling in Antarctica can hopefully improve estimates of $CO_2$ across the MPT.

There is evidence of a gradual deep ocean cooling since the onset of northern Hemisphere glaciation $2.7$ Million years ago \citep{lisiecki05}. How much of this cooling occurred across the MPT is not known, however. The reconstruction of deep water temperatures by \citep{elderfield12} indicates a gradual cooling in the mean from $-1300$ kyr until present, but also a puzzling warming from $-1500$ kyr to $-1300$ kyr. Therefore, glacial cycle length does not appear to have a direct relation with mean deep ocean temperature. However, it may be that sea surface temperatures in the vicinity of major ice sheets are more relevant for glacial dynamics. If so, detailed and reliable reconstruction of such temperatures is necessary to evaluate whether they act as ramped climate variables in RFL.

Another slowly varying parameter could be the erosion of regolith. According to this hypothesis soft material under ice sheets eroded throughout the Pleistocene, enabling them to grow larger before collapsing \cite{clark98}. The hypothesis is difficult to test empirically, however.

In addition to the candidate ramping climate variables mentioned, there may be others that are relevant for the MPT. RFL motivates the search for other such climate variables. These might not only be relevant for RFL, but for any mechanism of the MPT invoking deterministic bifurcation.

}

\section{Criteria for RFL}
\edtg{
Having demonstrated RFL in H07 and Paillard and Parennin 2004, we ask in which models RFL is most likely to be relevant.

We expect RFL in all models similar to H07, that is, models with a critical threshold of deglaciation (explicit or not), two intrinsic growth and decay states, additive forcing, and a climate variable that naturally controls the internal period.

Furthermore, RFL is facilitated by dynamics focussed on a single strongly attracting limit cycle. This is because solutions to the system $f$ with ramped parameters then track frozen solutions of $f_\tau$ well, and because it is difficult for perturbations to bring solutions away from the neighbourhood of the attractor.

Crucially, a model using RFL needs a parameter that can increase the internal period by $100$ kyr. The models in e.g. \citep{letreut83} and \citep{maasch90} are therefore difficult to reconcile with RFL since the internal periods are on the order of $10$ and $100$ kyr respectively, and do not change much within the physical range of model parameters.

Lastly, we note that e.g. excitable systems and dissipative resonant oscillators \citep{crucifix12} also can undergo a rapid change in durations due to frequency locking related phenomena, although they are not self-sustained oscillators. Self-sustained oscillators are distinguished by having an internal period $T_o(R)$ through which we can define Arnold tongue diagrams and Devil's staircases; for non-self-sustained oscillators we have to define these through parameters $R$ directly. Furthermore, it has been argued that the term frequency locking should be restricted to self-sustained oscillators \citep{pikovsky01,marchionne18}, why it makes sense to define RFL for self-sustained oscillators only.
}

\section{Conclusions}

\edt{
The glacial cycles did not enter a stationary $100$ kyr world at the MPT; instead, durations between glacial terminations shifted abruptly from approximately ${\sim}40$ to ${\sim}80$ kyr around $-1200$ kyr, followed by a gradual increase (Fig. \ref{fig:benthicdata}). The dynamical mechanism Ramping with Frequency Locking (RFL) naturally explains this progression of durations. As the internal period of a model glacial cycle model increases gradually, frequency locking to insolation variations causes the durations between glacial terminations to increase sometimes abruptly and sometimes gradually. 
}

The RFL mechanism is rather general and explains the behaviour of a range of models describing glacial cycles and the MPT \citep{feng15,huybers07,crucifix11,paillard98,paillard04,tzedakis17,ashkenazy06,mitsui15}.
 
Here we described how RFL can be understood in terms of a dynamical system $f$ and a frozen system $f_\tau$ with parameters $R(t)$ fixed at times $t=\tau$. The average duration $\overline{D}_\tau$ defined for $f_\tau$ provides some information about single durations in solutions $x(t)$ to $f$ around $t=\tau$, but since $\overline{D}_\tau$ is defined asymptotically, one must interpret solutions to $f$ in terms of $f_\tau$ with care.

Model behaviour can be understood from considering parameter paths through Arnold tongue diagrams and corresponding Devil's staircases (Fig. \ref{fig:solutionsavgdurations}, \ref{fig:arnoldtongues} and \ref{fig:arnoldtonguesqueezecomparison}). These diagrams as functions of frozen time $\tau$ depend on
\begin{itemize}
\item the change in average duration $\overline{D}_\tau(T_o)$ as function of $T_o$,
\item the change in internal period $T_o(R)$ as function of $R$,
\item the change in parameters $R(\tau)$ as function of time $\tau$, and
\item the amplitude $A$ of the forcing.
\end{itemize}

This decomposition clarifies different ways in which the average duration can change abruptly in models of the class $f$. For instance, the abruptness of the change in $\overline{D}_\tau$ can be adjusted either by changing the forcing amplitude $A$ or the ramping of $R(t)$. While the effects of changing $A$ or the ramping of $R(t)$ are typically easy to guess, we are not aware of any general rules dictating the widths of particular Arnold tongues. Such understanding may be researched further.

\edt{
RFL is relevant also for multi-frequency astronomical forcing. Multi-frequency forcing tends to make Devil's staircases less abrupt, but durations can still increase rapidly when a model parameter is slowly ramped.
}

The RFL mechanism provides an explanation for the MPT without the climate system entering a new mode of operation. A shift from ${\sim}40$ kyr long to ${\sim}80$ kyr long cycles due to frequency locking to \edt{obliquity and precession}, is consistent with data (Fig. \ref{fig:benthicdata}), and is used in models \citep{paillard04,huybers07}. This warrants further study of frequency locking characteristics of models throughout the model hierarchy\edt{, as well as a search for gradually increasing climate parameters}. Some models use a rapidly ramped parameter to accelerate the increase in durations between glacial terminations at the MPT, but such a ramping begs for justification that a model relying solely on frequency locking does not require.

\section*{Acknowledgements}

We thank Peter Ashwin for valuable discussions.

This research has been funded by the European Union's Horizon 2020 innovation and research programme for the ITN CRIT\-ICS under the Marie Sk\l{}odow\-ska-Curie grant agreement No. 643073.

\appendix

\section{Kendall's tau}
\label{sec:kendalltau}

Kendall's tau, here denoted $\tau_K$, when testing for monotonicity of a sequence $\{D_i\}_{i=1}^n$ is defined as
\begin{linenomath*}
\begin{equation}
\tau_K = \frac{n_c - n_d}{\sqrt{n_0^2 - n_0n_1}},
\end{equation}
\end{linenomath*}
where $n_c - n_d= \sum_{i<j} \sign{(D_j - D_i)}$ is the number of pairs $(D_i,D_j)$ that are ordered ($D_j>D_i$) minus the number that is disordered, $n_0 = n(n-1)/2$ is the total number of pairs and $n_1 = \sum_k t_k(t_k -1)/2$ is the sum of the number of tied elements $t_k$ in the $k$:th group of tied elements. For example, the sequence $\{1,2,2\}$ has two ordered pairs $(1,2)$ and $(1,2)$, zero disordered pairs, and one tied pair $(2,2)$. Hence, $k=1$ such that $n_c - n_d=2$, $n_0 = 3$ and $n_1=1$, giving $\tau_K = 2/\sqrt{6} \approx 0.82$.

\section{Wavelets}
\label{sec:wavelets}
Wavelet spectra are estimated with the MATLAB\textregistered function \texttt{cwt}, using Morlet basis functions with bandwidth parameter $\omega_0=6$ \citep{torrence98}. Contours show wavelet amplitude (square root of variance) relative to the maximum, incremented in evenly spaced percentage units. The cone of influence marks the $e$-folding time of the amplitude of a discontinuity at the edge of the time interval. Inside the cone of influence edge effects are negligible \citep{torrence98}.

\section{Glacial terminations in LR04}
\label{sec:glacialterminations}
Major glacial terminations in the LR04 stack (Fig.~\ref{fig:benthicdata}) are identified at times $t = $-[$1948$, $1900$, $1863$, $1795$, $1748$, $1708$, $1655$, $1575$, $1535$, $1496$, $1456$, $1412$, $1372$, $1336$, $1290$, $1248$, $1198$, $1126$, $1038$, $964$, $876$, $794$, $718$, $630$, $536$, $434$, $341$, $252$, $140$, $18]$ kyr. We assume conservatively an age model uncertainty with constant standard deviation $6$ kyr over the past $2000$ kyr \citep{lisiecki05}, which gives a standard deviation $\sqrt{2}\cdot 6$kyr on the durations between terminations, assuming somewhat wrongly that errors are independent and normally distributed.

{\RaggedRight
\bibliographystyle{spbasic}
%\bibliography{references}

\begin{thebibliography}{56}
\providecommand{\natexlab}[1]{#1}
\providecommand{\url}[1]{{#1}}
\providecommand{\urlprefix}{URL }
\expandafter\ifx\csname urlstyle\endcsname\relax
  \providecommand{\doi}[1]{DOI~\discretionary{}{}{}#1}\else
  \providecommand{\doi}{DOI~\discretionary{}{}{}\begingroup
  \urlstyle{rm}\Url}\fi
\providecommand{\eprint}[2][]{\url{#2}}

\bibitem[{Ashkenazy(2006)}]{ashkenazy06}
Ashkenazy Y (2006) The role of phase locking in a simple model for glacial
  dynamics. Climate Dynamics 27:421--431, \doi{10.1007/s00382-006-0145-5}

\bibitem[{Ashkenazy and Tziperman(2004)}]{ashkenazy04}
Ashkenazy Y, Tziperman E (2004) Are the 41 kyr glacial oscillations a linear
  response to {M}ilankovich forcing? Quaternary Science Reviews 23:1879--1890,
  \doi{10.1016/j.quascirev.2004.04.008}

\bibitem[{Ashwin and Ditlevsen(2015)}]{ashwin15}
Ashwin P, Ditlevsen P (2015) The middle {P}leistocene transition as a generic
  bifurcation on a slow manifold. Climate Dynamics 24,
  \doi{10.1007/s00382-015-2501-9}

\bibitem[{Ashwin et~al.(2018)Ashwin, Camp, and von~der Heydt}]{ashwin18}
Ashwin P, Camp CD, von~der Heydt AS (2018) Chaotic and non-chaotic response to
  quasiperiodic forcing: limits to predictability of ice ages paced by
  milankovitch forcing. Dynamics and Statistics of the Climate System 1(20),
  \doi{10.1093/climsys/dzy002}

\bibitem[{Baer et~al.(1989)Baer, Ernaux, and Rinzel}]{baer89}
Baer S, Ernaux T, Rinzel J (1989) The slow passage through a hopf bifurcation:
  Delay, memory effects, and resonance. {SIAM} {J}ournal on {A}pplied
  {M}athematics 49(1):55--71, \doi{10.1137/0149003}

\bibitem[{Berger(1978)}]{berger78}
Berger AL (1978) Long-{T}erm {V}ariations of {D}aily {I}nsolation and
  {Q}uaternary {C}limatic {C}hanges. Journal of the {A}tmospheric {S}ciences
  35, \doi{10.1175/1520-0469(1978)035<2362:LTVODI>2.0.CO;2}

\bibitem[{Cartwright and Littlewood(1945)}]{cartwright45}
Cartwright M, Littlewood J (1945) On non-linear differential equations of the
  second order. Journal of the {L}ondon {M}athematical {S}ociety
  1-20(3):180--189, \doi{10.1112/jlms/s1-20.3.180}

\bibitem[{Clark and Pollard(1998)}]{clark98}
Clark PU, Pollard D (1998) Origin of the middle {P}leistocene transition by ice
  sheet erosion of regolith. Paleoceanography 13(1):1--9,
  \doi{10.1029/97PA02660}

\bibitem[{Clark et~al.(2006)Clark, Archer, Pollard, Blum, Rial, Brovkin, Mix,
  Pisias, and Roy}]{clark06}
Clark PU, Archer D, Pollard D, Blum JD, Rial JA, Brovkin V, Mix AC, Pisias NG,
  Roy M (2006) The middle {P}leistocene transition: characteristics,
  mechanisms, and implications for the long-term changes in atmospheric
  {pCO$_2$}. Quaternary {S}cience {R}eviews 25:3150--3184,
  \doi{10.1016/j.quascirev.2006.07.008}

\bibitem[{Crucifix(2012)}]{crucifix12}
Crucifix M (2012) Oscillators and relaxation phenomena in pleistocene climate
  theory. Philosophical transactions of the {R}oyal {S}ociety of {L}ondon A
  370(1962):1140--1165, \doi{10.1098/rsta.2011.0315}

\bibitem[{Crucifix(2013)}]{crucifix13}
Crucifix M (2013) Why could ice ages be unpredictable? Climate of the {P}ast
  9:2253--2267, \doi{10.5194/cp-9-2253-2013}

\bibitem[{Crucifix et~al.(2011)Crucifix, Lenoir, and de~Saedeleer}]{crucifix11}
Crucifix M, Lenoir G, de~Saedeleer B (2011) The mid-{P}leistocene transition
  and slow fast dynamics. In: EGU2011-3629-1, EGU General Assembly 2011,
  Geophysical Research Abstracts, vol~13, poster

\bibitem[{Daruka and Ditlevsen(2015)}]{daruka15}
Daruka I, Ditlevsen PD (2015) A conceptual model for glacial cycles and the
  middle pleistocene transition. Climate Dynamics 46:29--40,
  \doi{10.1007/s00382-015-2564-7}

\bibitem[{Ditlevsen(2009)}]{ditlevsen09}
Ditlevsen PD (2009) Bifurcation structure and noise-assisted transitions in the
  pleistocene glacial cycles. Paleoceanography 24, \doi{10.1029/2008PA001673}

\bibitem[{Do and Lopez(2012)}]{do12}
Do Y, Lopez JM (2012) Slow passage through multiple bifurcation points.
  American Institute of Mathematical Sciences 18(1):95--107,
  \doi{10.3934/dcdsb.2013.18.95}

\bibitem[{Elderfield et~al.(2012)Elderfield, Ferretti, Greaves, Crowhurst,
  McCave, Hodell, and Piotrowski}]{elderfield12}
Elderfield H, Ferretti P, Greaves M, Crowhurst S, McCave IN, Hodell D,
  Piotrowski AM (2012) Evolution of ocean temperature and ice volume through
  the {M}id-{P}leistocene climate transition. Science 337(6095):704--709,
  \doi{10.1126/science.1221294}

\bibitem[{{{EPICA community members}}(2004)}]{epica04}
{{EPICA community members}} (2004) Eight glacial cycles from an {A}ntarctic ice
  core. Nature 429:623--628, \doi{10.1038/nature02599}

\bibitem[{Feng and Bailer-Jones(2015)}]{feng15}
Feng F, Bailer-Jones CAL (2015) Obliquity and precession as pacemakers of
  pleistocene deglaciations. Quaternary {S}cience {R}eviews 122:166--179,
  \doi{10.1016/j.quascirev.2015.05.006}

\bibitem[{Gildor and Tziperman(2000)}]{gildor00}
Gildor H, Tziperman E (2000) Sea ice as the glacial cycles’ climate switch:
  role of seasonal and orbital forcing. Paleoceanography 15(6):605--615,
  \doi{doi.org/10.1029/1999PA000461}

\bibitem[{Glass and Mackey(1979)}]{glass79}
Glass L, Mackey MC (1979) A simple model for phase locking of biological
  oscillators. Journal of {M}athematical {B}iology 7:339--352,
  \doi{10.1007/BF00275153}

\bibitem[{Grebogi et~al.(1984)Grebogi, Ott, Pelican, and Yorke}]{grebogi84}
Grebogi C, Ott E, Pelican S, Yorke JA (1984) Strange attractors that are not
  chaotic. Physica D 13:261--268, \doi{10.1016/0167-2789(84)90282-3}

\bibitem[{Guckenheimer et~al.(2003)Guckenheimer, Hoffman, and
  Weckesser}]{guckenheimer03}
Guckenheimer J, Hoffman K, Weckesser W (2003) The forced van der pol equation
  i: The slow flow and its bifurcations. {SIAM} {J}ournal on {A}pplied
  {D}ynamical {S}ystems 2(1):1–35, \doi{10.1137/S1111111102404738}

\bibitem[{Huybers(2006)}]{huybers06}
Huybers P (2006) Early pleistocene glacial cycles and the integrated summer
  insolation forcing. Science 313:508--510, \doi{10.1126/science.1125249}

\bibitem[{Huybers(2007)}]{huybers07}
Huybers P (2007) Glacial variability over the last two million years: an
  extended depth-derived agemodel, continuous obliquity pacing, and the
  {P}leistocene progression. Quaternary {S}cience {R}eviews 26:37--55,
  \doi{10.1016/j.quascirev.2006.07.013}

\bibitem[{Huybers(2009)}]{huybers09}
Huybers P (2009) Pleistocene glacial variability as a chaotic response to
  obliquity forcing. Climate of the Past 5:481--488,
  \doi{10.5194/cp-5-481-2009}

\bibitem[{Huybers(2011)}]{huybers11}
Huybers P (2011) Combined obliquity and precession pacing of late pleistocene
  deglaciation. Nature 480:229--231, \doi{10.1038/nature10626}

\bibitem[{Huybers and Langmuir(2017)}]{huybers17}
Huybers P, Langmuir CH (2017) Delayed {CO$_2$} emissions from mid-ocean ridge
  volcanism as a possible cause of late-pleistocene glacial cycles. {E}arth and
  {P}lanetary {S}cience {L}etters 457:238--249, \doi{10.1016/j.epsl.2016.09.0}

\bibitem[{Hönisch et~al.(2009)Hönisch, Hemming, Archer, Siddall, and
  McManus}]{honisch09}
Hönisch B, Hemming G, Archer D, Siddall M, McManus JF (2009) Atmospheric
  carbon dioxide concentration across the {M}id-{P}leistocene transition.
  Science 324(5934):1551--1554, \doi{10.1126/science.1171477}

\bibitem[{Imbrie and Imbrie(1980)}]{imbrie80}
Imbrie J, Imbrie JZ (1980) Modeling the climatic response to orbital
  variations. Science 207:943--953, \doi{10.1126/science.207.4434.943}

\bibitem[{Imbrie and Imbrie(1979)}]{imbrie79}
Imbrie J, Imbrie KP (1979) Ice ages: solving the mystery, 1st edn. MacMillan,
  London

\bibitem[{Imbrie et~al.(2011)Imbrie, Imbrie-Moore, and Lisiecki}]{imbrie11}
Imbrie JZ, Imbrie-Moore A, Lisiecki L (2011) A phase-space model for
  {P}leistocene ice volume. Earth and {P}lanetary {S}cience {L}etters
  307:94--102, \doi{10.1016/j.epsl.2011.04.018}

\bibitem[{Kendall(1955)}]{kendall55}
Kendall M (1955) Rank correlation methods, 2nd edn. Hafner Publishing Co.,
  Oxford, {E}ngland

\bibitem[{Laskar et~al.(2004)Laskar, Robutel, Joutel, Gastineau, Correia, and
  Levrard}]{laskar04}
Laskar J, Robutel P, Joutel F, Gastineau M, Correia ACM, Levrard B (2004) A
  long-term numerical solution for the insolation quantities of the earth.
  Astronomy and {A}strophysics 428:261--285, \doi{10.1051/0004-6361:20041335}

\bibitem[{{Le Treut} and Ghil(1983)}]{letreut83}
{Le Treut} H, Ghil M (1983) Orbital forcing, climatic interactions, and
  glaciation cycles. Journal of Geophysical Research 88(C9):5167--5190,
  \doi{10.1029/JC088iC09p05167}

\bibitem[{Levi(1990)}]{levi90}
Levi M (1990) A period-adding phenomenon. {SIAM} {J}ournal on {A}pplied
  {M}athematics 50(4):943–955, \doi{10.1137/0150058}

\bibitem[{Lisiecki and Raymo(2005)}]{lisiecki05}
Lisiecki LE, Raymo ME (2005) A {P}lioene-{P}leistocene stack of 57 globally
  distributed benthic {$\delta^{18}O$} records. Paleoceanography 20:437--440,
  \doi{10.1029/2004PA001071}

\bibitem[{Maasch and Salzman(1990)}]{maasch90}
Maasch KA, Salzman B (1990) {A} {L}ow-{O}rder {D}ynamical {M}odel of {G}lobal
  {C}limatic {V}ariability {O}ver the {F}ull {P}leistocene. {J}ournal of
  {G}eophysical {R}esearch 95(D2):1955--1963, \doi{10.1029/JD095iD02p01955}

\bibitem[{Mann(1945)}]{mann45}
Mann HB (1945) Nonparametric tests against trend. Econometrica 13(3):245–259,
  \doi{10.2307/1907187}

\bibitem[{Marchionne et~al.(2018)Marchionne, Ditlevsen, and
  Wieczorek}]{marchionne18}
Marchionne A, Ditlevsen P, Wieczorek S (2018) Is the astronomical forcing a
  reliable and unique pacemaker for climate? {A} conceptual study. Physica {D}
  380-381:8--16, \doi{10.1016/j.physd.2018.05.004}

\bibitem[{Mitsui et~al.(2015)Mitsui, Crucifix, and Aihara}]{mitsui15}
Mitsui T, Crucifix M, Aihara K (2015) {B}ifurcations and strange nonchaotic
  attractors in a phase oscillator model of glacial–interglacial cycles.
  {P}hysica {D} 306:25--33, \doi{10.1016/j.physd.2015.05.007}

\bibitem[{Mudelsee and Schulz(1997)}]{mudelsee97}
Mudelsee M, Schulz M (1997) The {M}id-{P}leistocene climate transtion: onset of
  100 ka cycles lags ice volume build-up by 280 ka. Earth and {P}lanetary
  {S}cience {L}etters 151:117--123, \doi{10.1016/S0012-821X(97)00114-3}

\bibitem[{Oerlemans(1980)}]{oerlemans80}
Oerlemans J (1980) Model experiments on the 100000-yr glacial cycle. Nature
  287:430--432, \doi{10.1038/287430a0}

\bibitem[{Omta et~al.(2015)Omta, Kooi, van Voorn, Rickaby, and
  Follows}]{omta16}
Omta AW, Kooi BW, van Voorn GAK, Rickaby REM, Follows MJ (2015) Inherent
  characteristics of sawtooth cycles can explain different glacial
  periodicities. Climate {D}ynamics 46:557--569,
  \doi{10.1007/s00382-015-2598-x}

\bibitem[{Paillard(1998)}]{paillard98}
Paillard D (1998) {T}he timing of {P}leistocene glaciations from a simple
  multiple-state climate model. {N}ature 391:378--381, \doi{10.1038/34891}

\bibitem[{Paillard and Parrenin(2004)}]{paillard04}
Paillard D, Parrenin F (2004) The {A}ntarctic ice sheet and the triggering of
  deglaciations. {E}arth and {P}lanetary {S}cience {L}etters 227:263--271,
  \doi{10.1016/j.epsl.2004.08.023}

\bibitem[{Parrenin and Paillard(2003)}]{parrenin03}
Parrenin F, Paillard D (2003) Amplitude and phase of glacial cycles from a
  conceptual model. Earth and Planetary Science Letters 214(1):243 -- 250,
  \doi{10.1016/S0012-821X(03)00363-7}

\bibitem[{Parrenin and Paillard(2012)}]{parrenin12}
Parrenin F, Paillard D (2012) Terminations vi and viii (530 and 720 kyr bp)
  tell us the importance of obliquity and precession in the triggering of
  deglaciations. Climate of the Past 8(6):2031--2037,
  \doi{10.5194/cp-8-2031-2012}

\bibitem[{Pikovsky et~al.(2001)Pikovsky, Rosenblum, and Kurths}]{pikovsky01}
Pikovsky A, Rosenblum M, Kurths J (2001) Synchronization: A universal
  phenomenon in the nonlinear sciences, 1st edn. Cambridge University Press,
  Cambridge

\bibitem[{van~der Pol and van~der Mark(1927)}]{vanderpol27}
van~der Pol B, van~der Mark J (1927) Frequency demultiplication. Nature
  120:363–364, \doi{10.1038/120363a0}

\bibitem[{Quinn et~al.(2018)Quinn, Sieber, von~der Heydt, and Lenton}]{quinn18}
Quinn C, Sieber J, von~der Heydt AS, Lenton TM (2018) The {M}id-{P}leistocene
  transition induced by delayed feedback and bistability. Dynamics and
  Statistics of the Climate System 3(1):1--17, \doi{10.1093/climsys/dzy005}

\bibitem[{de~Saedeleer et~al.(2013)de~Saedeleer, Crucifix, and
  Wieczorek}]{desaedeleer13}
de~Saedeleer B, Crucifix M, Wieczorek S (2013) Is the astronomical forcing a
  reliable and unique pacemaker for climate? {A} conceptual study. Climate
  {D}ynamics 40:273--294, \doi{10.1007/s00382-012-1316-1}

\bibitem[{Salzman and Verbitsky(1993)}]{salzman93}
Salzman B, Verbitsky MY (1993) Multiple instabilities and modes of glacial
  rhythmicity in the {P}lio-{P}leistocene: a general theory of late {C}enozoic
  climatic change. {C}limate {D}ynamics 9:1--15, \doi{10.1007/BF00208010}

\bibitem[{Torrence and Compo(1998)}]{torrence98}
Torrence C, Compo GP (1998) A practical guide to wavelet analysis. {B}ulletin
  of the {A}merican {M}eteorologial {S}ociety 79(1):61--78,
  \doi{10.2307/1907187}

\bibitem[{Tzedakis et~al.(2017)Tzedakis, Crucifix, Mitsui, and
  Wolff}]{tzedakis17}
Tzedakis PC, Crucifix M, Mitsui T, Wolff EW (2017) A simple rule to determine
  which insolation cycles lead to interglacials. Nature 542:427--432,
  \doi{10.1038/nature21364}

\bibitem[{Tziperman and Gildor(2003)}]{tziperman03}
Tziperman E, Gildor H (2003) On the mid-{P}leistocene transition to 100-kyr
  glacial cycles and the asymmetry between glaciation and deglaciation times.
  Paleoceanography 18(1):1--8, \doi{10.1029/2001PA000627}

\bibitem[{Tziperman et~al.(2006)Tziperman, Raymo, Huybers, and
  Wunsch}]{tziperman06}
Tziperman E, Raymo M, Huybers P, Wunsch C (2006) Consequences of pacing the
  pleistocene 100 kyr ice ages by nonlinear phase locking to milankovitch
  forcing. Paleoceanography 21:1--11, \doi{10.1029/2005PA001241}

\end{thebibliography}

}

\end{document}